\documentclass[journal=jacsat,manuscript=article,layout=twocolumn]{achemso}

\usepackage{geometry}

\setkeys{acs}{articletitle=true}
\usepackage{lettrine}

\usepackage[version=3]{mhchem} % Formula subscripts using \ce{}
\usepackage[T1]{fontenc}       % Use modern font encodings

\usepackage[fontsize=10pt]{scrextend}

\usepackage{newtxtext}
\usepackage{bm} 
\usepackage[labelfont=bf]{caption} % bold ``Figure''

\usepackage[hyperfootnotes=false]{hyperref}	 % make refs, cites, links in color
\hypersetup{
  colorlinks,
  citecolor=blue,
  linkcolor=blue,
  urlcolor=blue
}

\usepackage{titlesec} % inline subsections
\titleformat{\subsection}[runin]
{\normalfont\bfseries}{\thesubsection{.}}{1em}{}[.]

\titleformat{\section}
{\normalfont\sffamily\large\bfseries}{\thesection{.}}{1em}{\MakeUppercase}

\usepackage[table]{xcolor}
\usepackage{graphicx}
\usepackage{amsmath,fullpage,amssymb}
\usepackage{graphics,epsfig}
\usepackage{makeidx}
\usepackage{gensymb}
\usepackage{siunitx}

\usepackage{soul}	% for strikethrough font

\def\ln{{\operatorname{ln}}}

\def\rmd{{\mathrm{d}}}

\def\rme{{\mathrm{e}}}

\def\Eq{Eq.}
\def\Eqs{Eqs.}

\def\Fig{\textcolor{blue}{Fig.}}
\def\Figs{Figs.}

\def\ie{{\em i.e.}}

\newcommand{\kB}{k_\textrm{B}}

\newcommand{\del}[1]{\textcolor{orange} {}}

\newcommand{\trm}[1]{{\textrm{#1}}}
\newcommand{\new}[1]{\textcolor{black} {#1}}

\SectionNumbersOn
\let\oldmaketitle\maketitle
\let\maketitle\relax

\title{\flushleft Apparent line tension induced by surface-active impurities}

\author{Fabio Staniscia}
\affiliation{\rm\small Jo\v{z}ef Stefan Institute, Jamova 39, 1000 Ljubljana, Slovenia}

\author{Matej Kandu\v{c}}
\affiliation{\rm\small Jo\v{z}ef Stefan Institute, Jamova 39, 1000 Ljubljana, Slovenia}
\email{matej.kanduc@ijs.si}

\begin{document}
\pagenumbering{arabic}
\noindent

\parindent=0cm
\setlength\arraycolsep{2pt}

\twocolumn[	% make wide abstract
\begin{@twocolumnfalse}
\oldmaketitle

\begin{abstract}\small
Line tension in wetting processes is of high scientific and technological relevance, but its understanding remains vague, mainly because of its difficult determination.
A widely used method to extract a line tension relies on the variation of a droplet's contact angle with the droplet's size.
Such an approach yields the apparent line tension, which is an effective parameter that factors in numerous contributions to the finite-size dependence, thus masking the actual line tension in terms of the excess free energy of the three-phase contact line.
Based on our recent computer simulation study, we investigate how small amounts of nonionic surfactants, such as surface-active impurities, contribute to the apparent line tension in aqueous droplets.
When depositing polydisperse droplets, their different surface-area-to-volume ratios can result in different final bulk concentrations of surfactants, different excess adsorptions to the interfaces, and, consequently, different contact angles. We show that already trace amounts of longer-chained surfactants \new{in a pre-contaminated liquid are enough to affect measurements of the apparent line tension}. Our analysis quantifies to what extent ``background'' impurities, inevitably present in all kinds of experimental settings, limit the resolution of line-tension measurements, which is crucial for avoiding data misinterpretation.
\vspace{5ex}
\end{abstract}
\end{@twocolumnfalse}]

\maketitle
\setlength\arraycolsep{2pt}
\small

\section{Introduction}

The research of wetting phenomena has a long and rich history, owing to its ubiquity in natural and engineering processes~\cite{de1985wetting, Bonn09}.
Its core lies in measurements of contact angles of sessile droplets on solid substrates, which is considered the gold standard in wettability characterization~\cite{liu2019improving}.
The starting point is provided by the classical Young equation
$\cos\theta_0 = (\gamma_\trm{sv}-\gamma_\trm{sw})/\gamma$, which relates the contact angle of a macroscopic  droplet, $\theta_0$, on a partially wettable surface to the surface--vapor ($\gamma_\trm{sv}$), surface--water ($\gamma_\trm{sw}$), and water--vapor ($\gamma$) surface tensions, in the case of water as the liquid.
 The Young equation, by which the contact angle is independent of the droplet size, has been widely used and confirmed for large, macroscopic droplets. 
A historically influential extension was introduced by Gibbs~\cite{gibbsLineTension}, who postulated the concept of line tension, $\tau$, as the excess free energy per unit length of the three-phase contact line of the droplet. Its implementation leads to the modified Young equation, also termed the Boruvka--Neumann equation~\cite{boruvka1977generalization, gaydos1987dependence}
\begin{equation}
\cos\theta=\cos\theta_{0}-\frac{\tau}{\gamma a}
\label{eq:cosvs1a}
\end{equation}
according to which the contact angle becomes dependent on the droplet's size.
Here, the second term on the right-hand side represents the correction due to the line tension to the macroscopic value, with $a$ being the droplet's base radius. 

The significance of line tension reaches far beyond the contact angles of sessile droplets. Line tension plays a crucial role in heterogeneous nucleation~\cite{joanny1986role, law1994theory, retter1996determination, lefevre2004intrusion}, vaporization~\cite{zhang2014influence}, droplet spreading and fragmentation~\cite{fan2006liquid, paneru2015liquid}, the stability of foams, films, filaments, droplets, and nanobubbles~\cite{aronson1994influence, rosso2004sign, brinkmann2005stability, guzzardi2006residual,mechkov2007contact, guillemot2012activated}, and even in hydrophobic interactions~\cite{sharma2012evaporation} to name but a few.
Therefore, understanding line tension is paramount for many phenomena in different scientific and technological fields.

The origin of line tension has been studied extensively and attributed to the imbalance of the intermolecular forces in the three-phase confluence region
\cite{indekeu1994line,amirfazli2004status, heim2013measurement}. Experiments of various liquids and substrates report the span of line tensions over seven orders of magnitude, ranging from $10^{-12}$ to $10^{-5}$~N and of either sign~\cite{drelich1992line, duncan1995correlation, pompe2000three, leelamanie2012drop, law2017line, matsubara2018finite, zhao2019resolving}.
On the contrary, most theoretical studies estimate line tension magnitudes near the lower limit of the experimental range~\cite{indekeu1992line, getta1998line, amirfazli2004status, schimmele2007conceptual}, leaving the larger values in experiments poorly understood.

Over time, \Eq~\ref{eq:cosvs1a} has become the standard for determining the line tension from the plots of $\cos\theta$ against the inverse base radius of sessile droplets in experimental and simulation studies~\cite{wang1999line, berg2010impact, werder2003water, zhang2014influence, zhang2018critical}.
Yet, it has become clear that there are additional contributions to $\cos\theta$ in wetting processes that scale inversely with radius and, with that, obscure the ``actual'' line tension as originally defined by Gibbs. The contributing physical effects can originate from the contact line itself (e.g., line tension stiffness, contact line pinning)~\cite{marmur2002line, rusanov2004line, schimmele2007conceptual, kanduc2017going, zhang2018contact} or from the interface and bulk regions of the droplet and have nothing to do with the three-phase contact line. Examples of those include the curvature-dependent surface tension~\cite{schimmele2007conceptual, kanduc2017going}, pressure-dependent surface tension~\cite{ward2008effect, schimmele2009line, tatyanenko2017comparable}, gravitational effects (for millimeter-sized droplets)~\cite{law2017line}, and adsorption of surface-active molecules---which is the topic of this paper. %In the case that the different contributions are independent, they can be summed up into the apparent line  tension~\cite{law2017line}.
Therefore, extracting the line tension from the fit of the modified Young equation (\Eq~\ref{eq:cosvs1a}) is today regarded as the {\it pseudo} or {\it apparent line tension}---an empirical parameter that lumps together all size-dependent contributions to the contact angle~\cite{drelich1992line, law2017line, kanduc2018generalized, zhang2018critical, iwamatsu2018generalized}. 
In that sense, the term is a misnomer as it does not necessarily have to do {\it only} with the three-phase contact line, but it merely characterizes deviations from the ``ideal'' Young case.
Despite laborious experimental and theoretical efforts in the past decades, line tension remains an elusive quantity up to this day.

% contamination
One of the widely recognized hurdles in contact angle measurements is contamination~\cite{Chang95, an2015wetting}.
It is, unfortunately, difficult to prove that measurements made in aqueous systems are free from impurities.
Impurities, ranging from surface-active molecules to insoluble organic materials, can be present in the liquid, on the substrate, or can come from contaminated air (i.e., airborne contamination)~\cite{ponce2016effects}.  Airborne hydrocarbons exist almost everywhere, even inside nanofabrication cleanrooms~\cite{smith1998analysis}.
For instance, it has been recognized that airborne contamination greatly impacts the wettability of some materials, including silica surfaces~\cite{iglauer2014contamination, saraji2014effects}, transition-metal semiconductors~\cite{chow2015wetting, gurarslan2016van}, and
graphene~\cite{li2013effect, kozbial2014understanding}, leading to the long-lasting controversy about its actual contact angle.

Impurities impact wetting in multiple ways, depending on their type.
For instance, insoluble impurities that strongly adsorb to the substrate form a static adlayer. Its effect can be tackled, for example, by Lifshitz-based models and density functional theory calculations of optical properties~\cite{zhou2018microscopic}.
On the contrary, if impurities are soluble surface-active molecules, their adsorption to droplet's interfaces modifies the contact angle via the Lucassen-Reynders equation~\cite{lucassen1963contact}. %Well below their critical aggregation concentration, the effect is linearly proportional to the impurity concentration.
In fact, not much attention has been devoted to the processes in which droplets end up with surfactants whose concentration systematically varies with the droplet's size.
%If the protocol of generating droplets lead to different surfactant concentration in different droplet's size, this engenders an apparent line tension contribution.

%In this study, we analyze how small amounts of surfactants modify the contact angle of a droplet.
In this study, we analyze two common protocols for depositing polydisperse sessile droplets that result in a systematic variation in the internal bulk concentration of surfactants with droplet size. The size variation leads to a size-dependent contact angle and hence engenders a contribution to the apparent line tension. We will demonstrate the outcomes based on the cases of linear alcohols, which provide simple insight into the role that the chain length plays in these processes.

\section{Adsorption of surfactants and the contact angle}
\begin{figure}[h]\begin{center}
\begin{minipage}[b]{0.38\textwidth}\begin{center}
\includegraphics[width=\textwidth]{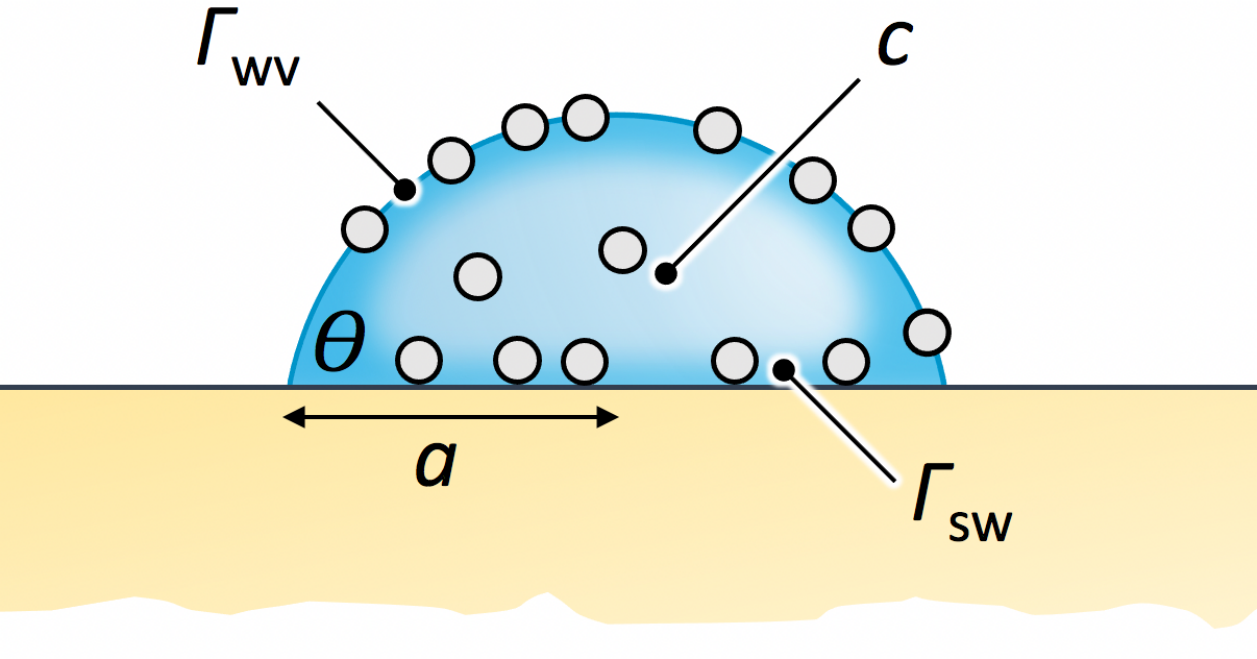} 
\end{center}\end{minipage}
\caption{Distribution of surfactants (grey circles) in a sessile droplet of contact angle $\theta$ and base radius $a$: Internal bulk concentration is denoted by $c$, and adsorptions to the water--vapor and substrate--water interfaces are denoted by $\Gamma_\trm{wv}$ and $\Gamma_\trm{sw}$, respectively. }
\label{fig:droplet}
\end{center}\end{figure}

When surface-active molecules are dissolved in a water droplet, some of them will adsorb to accessible interfaces as schematically shown in \Fig~\ref{fig:droplet}.
At sufficiently low concentrations, the adsorption is linearly proportional to bulk concentration $c$, well inside the droplet.
Thus, for the adsorption (characterized as the surface excess) to the water--vapor, substrate--water, and substrate-vapor interfaces of the droplet, we write respectively
\begin{eqnarray}
\Gamma_\trm{wv}&=&K_\trm{v} c\label{eq:Gammawv}\\
\Gamma_\trm{sw}&=&K_\trm{s} c\label{eq:Gammasw}\\
\Gamma_\trm{sv}&=&0\label{eq:Gammasv}
\end{eqnarray}
where $K_\trm{v}$ and $K_\trm{s}$ are the {\it adsorption coefficients} to the water--vapor and substrate--water interface, respectively.
Here, we assume no adsorption at the bare substrate--vapor interface, which is the case for well-soluble surfactants~\cite{staniscia2022tuning}.
%For higher concentrations, the trend starts deviating from linearity, typically in a saturation-like behavior.

\new{Nonionic} organic surfactants are driven to nonpolar interfaces by the hydrophobic effect, owing to their nonpolar tails, typically alkyl chains made up of methyl groups (--CH$_2$--).
We will base our study on linear primary alcohols, in which the hydroxy group ($-$OH) is bonded to an alkyl chain of a variable length. For brevity, we will term them by the number of carbon atoms (\ie, the chain length) as C1, C2, etc.
Such a homologous series of linear alcohols is a simple and very insightful model system to study the surfactant effect based on the chain length.

\begin{figure}[h]\begin{center}
\begin{minipage}[b]{0.35\textwidth}\begin{center}
\includegraphics[width=\textwidth]{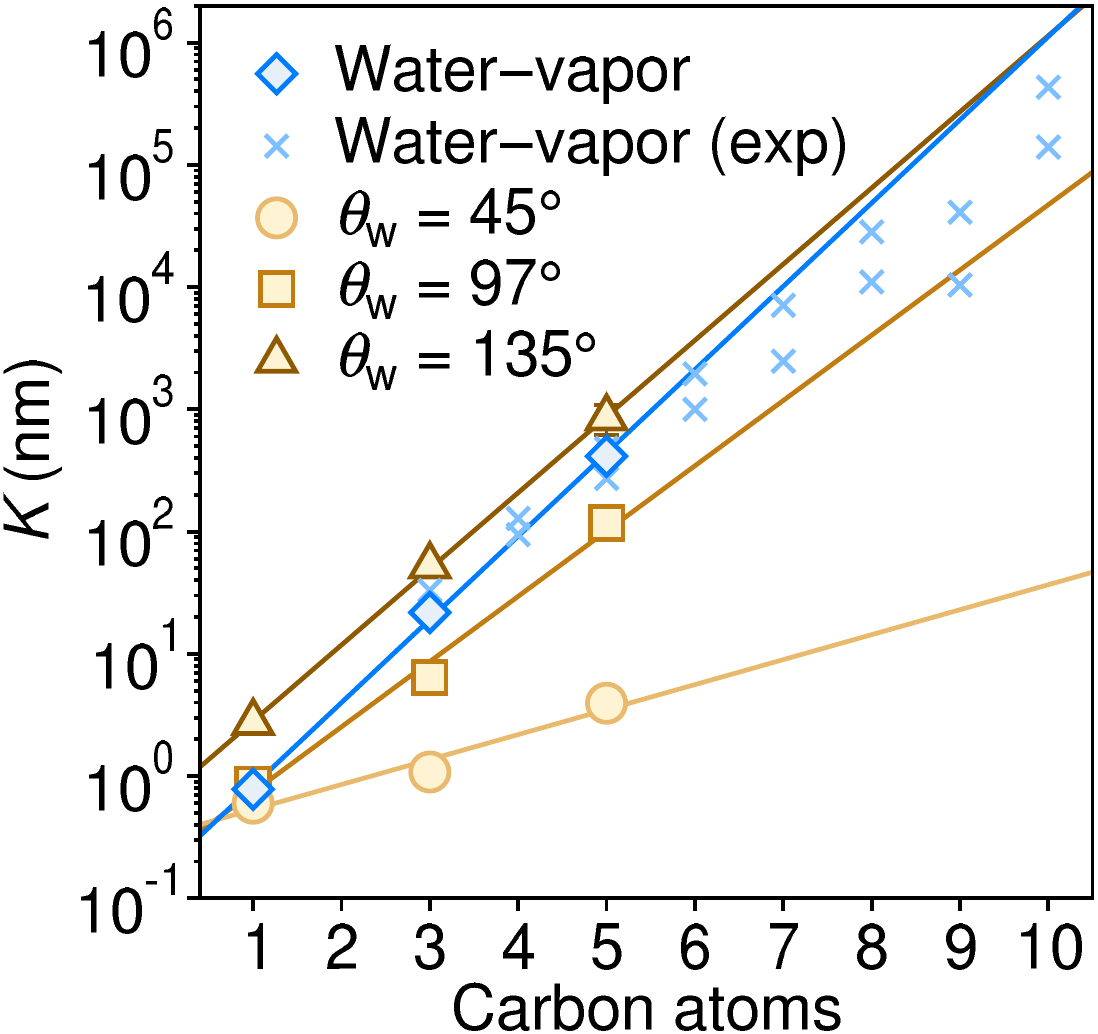} 
\end{center}\end{minipage}
\caption{Adsorption coefficients to water--vapor ($K_\trm{v}$) and substrate--water interfaces ($K_\trm{s}$) with different contact angles $\theta_\trm{w}$ for $n$-alcohols versus the number of alkyl carbon atoms. 
The MD simulation data are taken from Ref.~\citenum{staniscia2022tuning}. For comparison, experimental data for the water--vapor interface taken from Bleys and Joos~\cite{Bleys85} are shown by blue crosses.
The solid lines are the fits of \Eqs~\ref{eq:Kv} and~\ref{eq:Ks} to the MD data points.
The fitting coefficients are
$b_\trm{v}=0.173$~nm and $\alpha_\trm{v}=1.569$ for the water--vapor interface and 
$b_\trm{s}=0.333$~nm, $0.216$~nm, $0.674$~nm and $\alpha_\trm{s}=0.471$~nm, $1.229$~nm, $1.434$~nm for the substrate--water interfaces with $\theta_\trm{w}=45^\circ$, $97^\circ$, $135^\circ$, respectively.
}
\label{fig:K}
\end{center}\end{figure}

In \Fig~\ref{fig:K} we show adsorption coefficients obtained
in our recent molecular dynamics (MD) simulation study~\cite{staniscia2022tuning} for three short-chained $n$-alcohols (bulky symbols): methanol (C1), propanol (C3), and pentanol (C5).
%(henceforth referred to as C-1)
The results are shown for the water--vapor interface as well as solid substrates with three different contact angles $\theta_\trm{w}$ (tuned by the polarity of the surface groups), featuring a hydrophilic surface ($\theta_\trm{w}=45^\circ$), intermediate surface ($\theta_\trm{w}=97^\circ$), and hydrophobic surface ($\theta_\trm{w}=135^\circ$). These contact angles correspond to a pure water droplet without surfactants.

Hydrophobic hydration is known for its volume-to-area scaling crossover with an increasing hydrophobe size~\cite{chandler2005interfaces}. However, extending the alkyl chain length makes both the volume as well as the surface area increase linearly. Therefore, the transfer free energies scale very well with the number of carbon atoms in their tails~\cite{staniscia2022tuning, singh2022hydration}, and consequently, the adsorption coefficients scale exponentially with the number of carbon atoms. The simplest mathematical model that captures the behavior of the adsorption coefficients with chain length is thus,
\begin{eqnarray}
K_\trm{v}&=&b_\trm{v}\,\rme^{\alpha_\trm{v} n_\trm{C}}\label{eq:Kv}\\
K_\trm{s}&=&b_\trm{s}\,\rme^{\alpha_\trm{s} n_\trm{C}}\label{eq:Ks}
\end{eqnarray}
where $n_\trm{C}$ is the number of carbon atoms.
The coefficients $b_i$ and $\alpha_i$ are model parameters, which we fit to the simulation data, as shown by the lines in \Fig~\ref{fig:K}. 
%This will turn out useful later when to extrapolate the trends for larger molecules.
The fitted exponential trend for the water--vapor interface agrees reasonably well with experimental data extracted from the collection of Bleys and Joos~\cite{Bleys85} (blue crosses).
%Even though the fit to MD data overestimates 

The adsorption of surfactants to both droplet interfaces (\Fig~\ref{fig:droplet}) lowers their surface tensions and changes the wetting contact angle. The change can be calculated from the Lucassen-Reynders equation~\cite{lucassen1963contact},
\begin{equation}
\frac{\rmd (\gamma\cos\theta)}{\rmd \gamma}=\frac{\Gamma_\trm{sv}-\Gamma_\trm{sw}}{\Gamma_\trm{wv}}
\label{eq:LR}
\end{equation}
Using the adsorption relations given by \Eqs~\ref{eq:Gammawv}--\ref{eq:Gammasv}, leads to (see Appendix~\ref{sec:app:theta} for derivation),
%\begin{equation}
%K_\trm{v}\cos\theta+K_\trm{s}=(K_\trm{v}\cos\theta_\trm{w}+K_\trm{s})\exp\left(\frac{\kB T %K_\trm{v} c}{\gamma}\right)
%\label{eq:dtheta1}
%\end{equation}
\begin{equation}
\frac{K_\trm{v}\cos\theta+K_\trm{s}}{K_\trm{v}\cos\theta_\trm{w}+K_\trm{s}}=\exp\left(\frac{\kB T K_\trm{v} c}{\gamma}\right)
\label{eq:dtheta1}
\end{equation}
where $\theta$ and $\theta_\trm{w}$ refer to the contact angles of a surfactant-laden and pure water droplet, respectively.
For small changes in the water contact angle around $\theta_\trm{w}$, \Eq~\ref{eq:dtheta1} simplifies into~\cite{staniscia2022tuning}
\begin{equation}
\theta(c)=\theta_\trm{w}-\kB T\,\frac{K_\trm{v}\cos\theta_\trm{w}+K_\trm{s}}{\gamma \sin\theta_\trm{w}}\,c
\label{eq:dtheta}
\end{equation}
Surfactants (featuring positive  $K_\trm{v}$ and $K_\trm{s}$) mostly cause the contact angle to decrease (which is strictly true for $\theta_\trm{w}<90^\circ$).
%In the linear regime, the contact angle changes linearly with the concentration of added surfactant. 
%We will comment on the range validity at the end of our discussion.

\section{Droplet deposition and contact-angle variation}

\begin{figure*}[h]\begin{center}
\begin{minipage}[b]{0.75\textwidth}\begin{center}
\includegraphics[width=\textwidth]{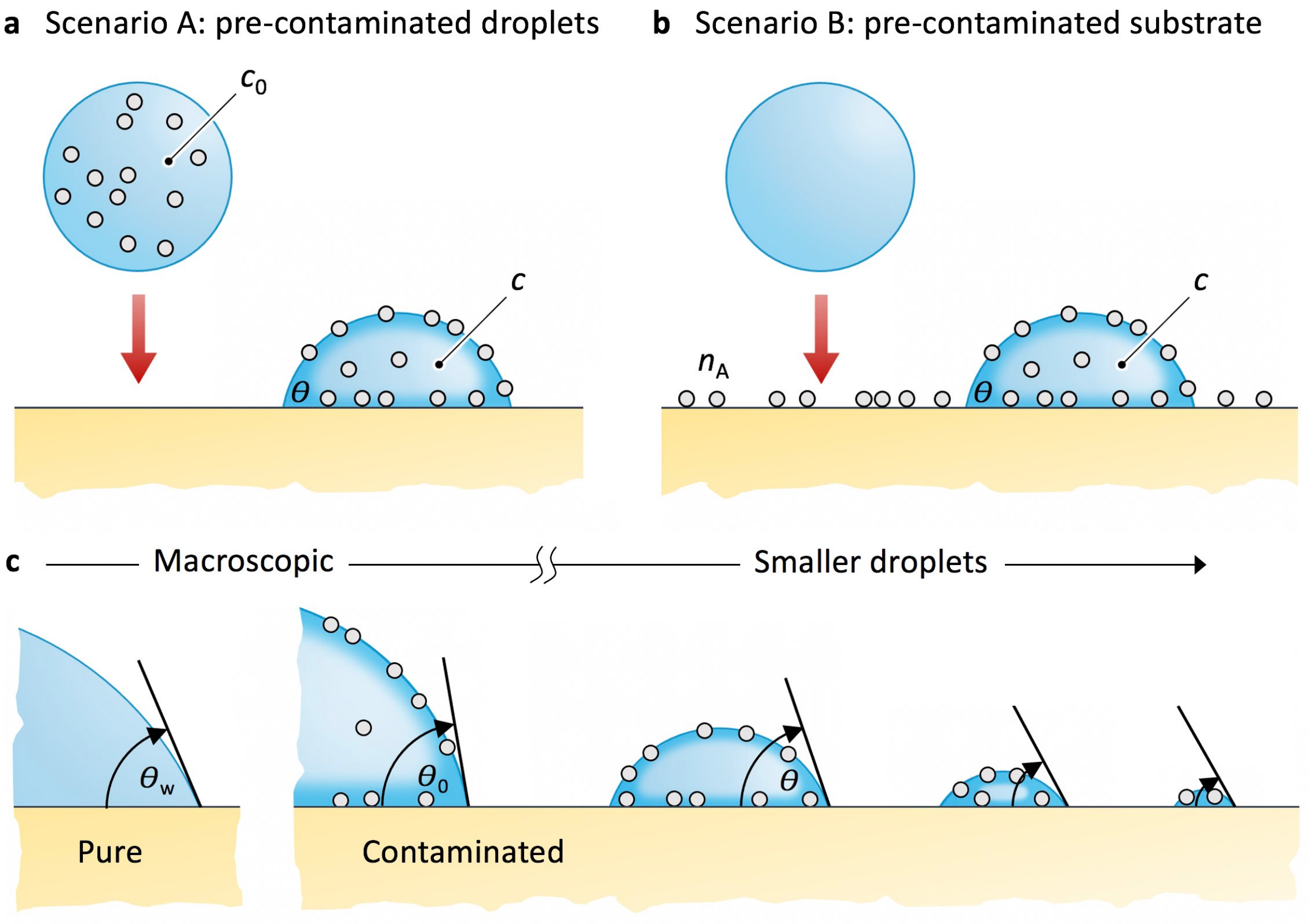}
\end{center}\end{minipage}
\caption{Different scenarios of droplet contamination that lead to size-dependent concentrations of surfactants. (a) Scenario~A: Pre-contaminated droplet with uniformly distributed contaminants of the total concentration that is independent of the droplet size. Upon adsorption on a clean substrate, the contaminants redistribute inside the droplet, resulting in a size-dependent bulk concentration $c$. 
(b) Scenario~B: A pure water droplet deposits on a pre-contaminated substrate with contaminants of areal density $n_A$. The contaminants hitherto adsorbed in the region of the droplet's base dissolve into the droplet's interior.
(c) Variation of the contact angle with droplets' size resulting from dissimilar contaminant concentration. The pure water contact angle is denoted by $\theta_\trm{w}$ and the macroscopic contact angle of a contaminant-laden droplet by $\theta_0$.
}
\label{fig:schematic}
\end{center}\end{figure*}

Based on \Eq~\ref{eq:dtheta}, the contact angle changes linearly with the surfactant concentration $c$ in the bulk interior of a droplet. Importantly, the change is not directly dependent on the droplet's size. 
However, depending on how the droplets get contaminated, the droplet size can influence the internal bulk concentration and, with that, indirectly the contact angle.
In this study, we will analyze two routes (named scenarios A and B) in which deposited polydisperse droplets end up with surfactant concentrations that systematically vary with their size.
In scenario A, all freshly generated droplets are pre-contaminated by the same total concentration of surfactants and are then deposited on a clean substrate (\Fig~\ref{fig:schematic}a).
In scenario B, pure water droplets are deposited onto a pre-contaminated substrate (\Fig~\ref{fig:schematic}b).

\subsection{Scenario A}
We first take a closer look at scenario A. 
As schematized in \Fig~\ref{fig:schematic}a, we here consider that all droplets at their formation contain uniformly distributed surfactants of concentration $c_0$, also referred to as {\em total concentration}. 

This scenario can be realized, for instance, in the atomization of a contaminated bulk liquid, which is rapidly dispersed into smaller droplets of various sizes (e.g., in the process of spraying and precipitation)~\cite{pompe2000three, wang2002evaluation}. 
Another very relevant protocol is droplet deposition by a syringe or pipette, as commonly used for creating larger droplets. 
While a droplet is forming at the tip of the syringe, surfactants first have to diffuse to the freshly formed air--water interface from the contaminated bulk interior. 
It is known that it takes several minutes for typical impurities to adsorb to a fresh air--water interface~\cite{ponce2016effects}. The adsorption time can be estimated as follows.
Assuming the adsorption to be a diffusion-controlled process, the time needed for the formation of the adsorbed layer can be roughly estimated from Sutherland's solution of the diffusion equation as
$\tau\approx K_\trm{v}^2/D$, where $D$ is the diffusion coefficient of surfactants in water~\cite{sutherland1952kinetics}. This estimate matches the time it takes for surfactants to diffuse over the distance $K_\trm{v}$, which corresponds to the slab thickness in the bulk solution that contains the same number of surfactants as adsorbed at the interface in equilibrium.
Diffusion coefficients only marginally depend on the chain length for short-chain alcohols and are on the order of $D\sim 1$~nm$^2/$ns~\cite{hao1996binary}, whereas the adsorption coefficients vary exponentially with chain length (\Eq~\ref{eq:Kv}).
With these assumptions, we obtain the estimated adsorption times of around $\tau\sim\,$0.1~ms for C5, 1~s for C8, and already 10 min for C10. 
This means that surfactants longer than around 8 or 9
carbon atoms cannot adsorb to the interface within typical times of droplet formations by a syringe. The syringe technique therefore classifies as scenario A for longer-chain surfactants.

After the contaminated droplets deposit on a clean substrate, the surfactants redistribute by diffusion inside each droplet to reach thermodynamic equilibrium, satisfying \Eqs~\ref{eq:Gammawv}--\ref{eq:Gammasv} (\Fig~\ref{fig:schematic}a).
By doing so, the internal bulk concentration in the droplets lowers from $c_0$ to $c$.
The conservation of the number of surfactants in the droplet after equilibration demands 
\begin{equation}
V c_0=V c+A_\trm{cap}\Gamma_\trm{wv}+A_\trm{base}\Gamma_\trm{sw}
\label{eq:conservationI}
\end{equation}
where the three terms on the right-hand side correspond to the total number of surfactants in the droplet's bulk, at the droplet's cap, and at its base area, respectively. See Appendix~\ref{sec:app:cap} for the geometrical expressions for the volume and surface areas.

This brings us to the following dependence of the bulk concentration on base radius
\begin{equation}
c=\frac{c_0}{1+a^*/a}
\label{eq:conc_bulk}
\end{equation}
where we have introduced the {\it partition radius} $a^*$ 
%\begin{equation}
%a^*=\frac{3\cot (\theta/2)}{2+\cos\theta}\left[2K_\trm{v}+(1+\cos\theta) K_\trm{s}\right]
%\end{equation}
\begin{equation}
a^*=6\, \frac{K_\trm{v}+ K_\trm{s}\cos^2(\theta_\trm{w}/2)}{\sin\theta_\trm{w}+\tan(\theta_\trm{w}/2)}
\label{eq:astar}
\end{equation}
It is not difficult to see from \Eq~\ref{eq:conc_bulk} that the ratio $a^*/a$ equals the ratio between the adsorbed and nonadsorbed surfactants in the droplet, thus, $a^*$ characterizes the length scale of surfactant partitioning.
%As we will see later on, the surfactants have the largest effect for the contact angle variations for droplets with the radius $a=a^*$.
%When the droplet's radius is equal the partition radius, $a=a^*$, the internal bulk concentration is half of the total concentration, $c=c_0/2$.
%It can also be seen from \Eq~\ref{eq:astar} that the partition radius is of the order of the adsorption coefficients, $a^*\sim \max\{K_\trm{v}, K_\trm{s}\}$.

In short, the variation of the internal concentration (\Eq~\ref{eq:conc_bulk}) has its origin in the fact that the surface-area-to-volume ratio varies with the droplet size. 
The depletion of the surfactants from the bulk is therefore more significant in smaller droplets.
For very large droplets, the depletion is small, and the bulk concentration remains unchanged, $c\approx c_0$. The macroscopic contact angle of surfactant-laden droplets is therefore $\theta_0\equiv \theta(c_0)$. See \Fig~\ref{fig:schematic}c for illustration.
Using \Eq~\ref{eq:dtheta}, we compute the change of the contact angle from a macroscopically large droplet $\Delta \theta(c)\equiv\theta(c)-\theta_0$, which is
\begin{equation}
\Delta\theta(c)=\kB T\,\frac{K_\trm{v}\cos\theta_\trm{w}+K_\trm{s}}{\gamma \sin\theta_\trm{w}}\, (c_0-c)
\end{equation}
and after expressing it in terms of base radius (using \Eq~\ref{eq:conc_bulk}),
\begin{equation}
\Delta\theta(a)=\kB T\,\frac{K_\trm{v}\cos\theta_\trm{w}+K_\trm{s}}{\gamma \sin\theta_\trm{w}}\, \frac{c_0}{1+a/a^*}
\label{eq:dtheta2}
\end{equation}

%In order to demonstrate the effect of surfactants on the contact angle, we use the data for homologous series of alcohols in \Fig~\ref{fig:K}.

\begin{figure*}\begin{center}
\begin{minipage}[b]{0.28\textwidth}\begin{center}
\includegraphics[width=\textwidth]{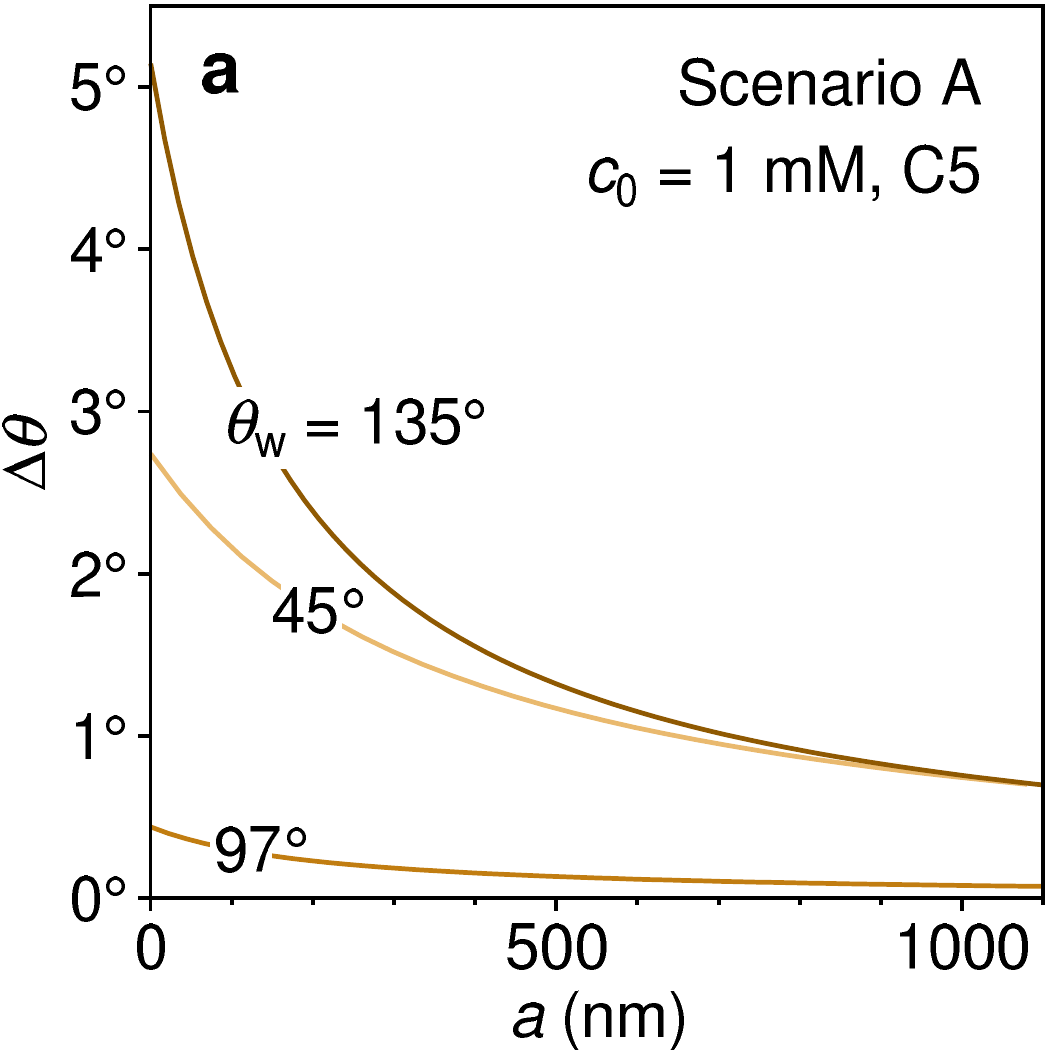}
\end{center}\end{minipage}\hspace{2ex}
\begin{minipage}[b]{0.3\textwidth}\begin{center}
\includegraphics[width=\textwidth]{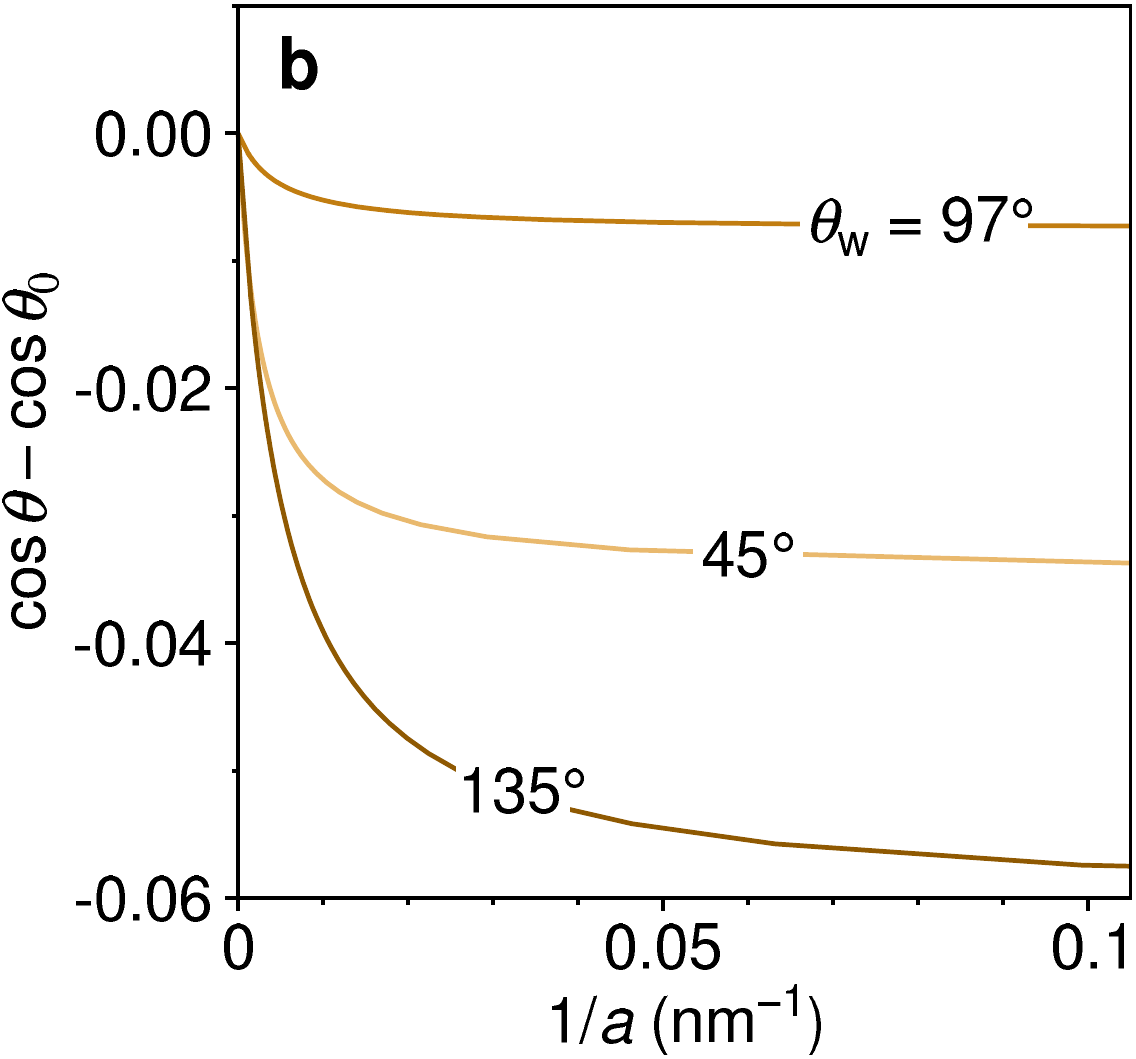}
\end{center}\end{minipage}\hspace{2ex}
\begin{minipage}[b]{0.29\textwidth}\begin{center}
\includegraphics[width=\textwidth]{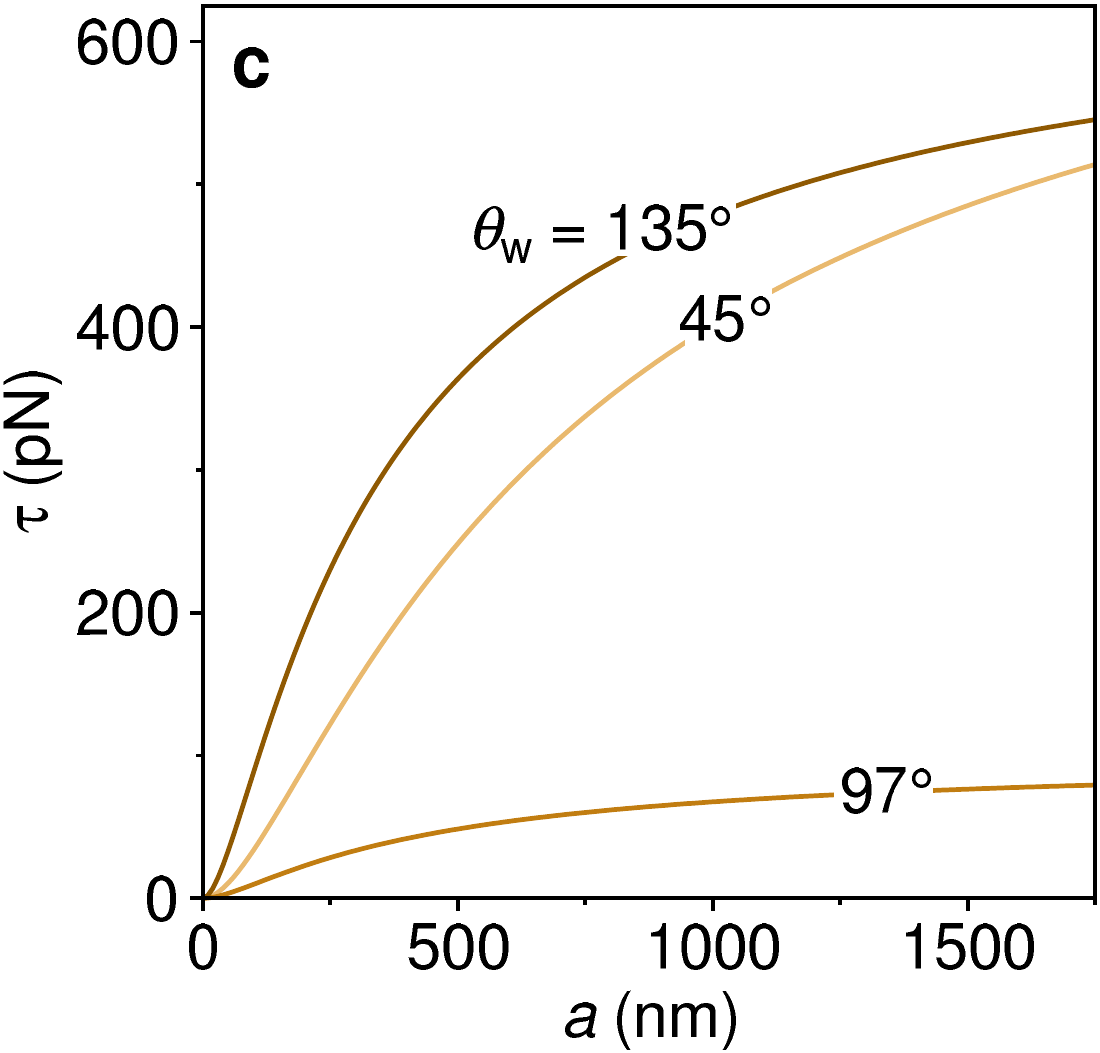}
\end{center}\end{minipage}
\caption{
(a) Variation of contact angles of aqueous droplets contaminated with pentanol of total concentration $c_0=1$ mM in scenario A on three different substrates as calculated with \Eq~\ref{eq:dtheta2}.
%Variation of the droplet's contact angle with its size in scenario A, calculated with \Eq~\ref{eq:dtheta2} for $c_0=1$ mM of pentanol and three different water contact angles of the substrate, $\theta_\trm{w}=45^\circ$, $97^\circ$, and $135^\circ$. 
(b) Cosine of the contact angle shifted by $\cos\theta_0$ versus $1/a$ of the cases from panel (a). 
(c) Apparent line tension calculated with \Eq~\ref{eq:taudef}.
}
\label{fig:theta}
\end{center}\end{figure*}

Using the above equation, we show in \Fig~\ref{fig:theta}a how the contact angle in scenario A varies as a function of base radius for droplets containing $c_0=1$ mM of pentanol (C5) on the three different substrates. As droplets get smaller, the surfactants deplete more from the droplets' interior, therefore, the contact angles \new{increase}.
The effect is nonmonotonic in terms of surface hydrophobicity: the variation is the smallest on the intermediate surface and larger on the hydrophilic and hydrophobic surfaces. This nonmonotonicity in wetting enhancement due to surfactants has been thoroughly investigated elsewhere~\cite{staniscia2022tuning}.

\subsection{Scenario B}
In scenario B, we consider deposition of pure water droplets on a substrate that is pre-contaminated with surfactants of areal density $n_A$, as shown in \Fig~\ref{fig:schematic}b. After a droplet is deposited, the number of $A_\trm{base}n_A$ surfactants grasped by the freshly deposited droplet dissolves into the droplet.
Let us mention that scenario B could possibly also be relevant to airborne contamination coming from the surrounding atmosphere, where the uptake of contaminants is proportional to the droplet's surface area.

The conservation of surfactants after equilibration in this case demands
\begin{equation}
A_\trm{base}n_A=V c+A_\trm{cap}\Gamma_\trm{wv}+A_\trm{base}\Gamma_\trm{sw}
\label{eq:conservationII}
\end{equation}
 We assume that after the droplet is deposited, no additional surfactants diffuse into the droplet, which is realized by slow enough diffusion of surfactants on the dry substrate, thereby justifying $\Gamma_\trm{sv}=0$ (\Eq~\ref{eq:Gammasv}).  

In this case, the internal bulk surfactant concentration depends on the droplet's size as
\begin{equation}
c=\frac{\tilde c_0}{1+a/a^*}
\end{equation}
where the partition radius $a^*$ is again given by \Eq~\ref{eq:astar}. Here, we have introduced 
\begin{equation}
\tilde c_0=\frac{n_A}{K_\trm{v}/\cos^2(\theta_\trm{w}/2)+K_\trm{s}}
\label{eq:ctilde}
\end{equation}
which corresponds to the maximal internal concentration, reached for small droplets $a\ll a^*$.
Contrary to scenario A, very large droplets (with a small surface-to-volume ratio) remain essentially pure, thus the macroscopic contact angle in this case is $\theta_0=\theta_\trm{w}$.
We again calculate the change of the contact angle from the macroscopic contact angle, $\Delta \theta(c)\equiv\theta(c)-\theta_\trm{w}$, which is
\begin{equation}
\Delta\theta(a)=-\kB T\,\frac{K_\trm{v}\cos\theta_\trm{w}+K_\trm{s}}{\gamma \sin\theta_\trm{w}}\, \frac{\tilde c_0}{1+a/a^*}
\label{eq:dtheta3}
\end{equation}
This result is similar to scenario A (\Eq~\ref{eq:dtheta2}) but with the opposite sign and $\tilde c_0$ instead of $c_0$. In this case, the contact angle grows with increasing size.
Because of the similarity in the functional dependence to scenario A, we will not make extra examples for this scenario.

\section{Apparent line tension}
In many experimental as well as simulation studies of sessile droplets, contact angle data are analyzed by plotting $\cos\theta$ against $1/a$. The slope is then used to determine the apparent line tension, as given by \Eq~\ref{eq:cosvs1a}.
Therefore, we plot the data from  Fig.~\ref{fig:theta}a as $\cos\theta-\cos\theta_0$ versus \ $1/a$ in Fig.~\ref{fig:theta}b.
For a better comparison, the data have been shifted by the extrapolated values of $\cos \theta_0$ so that all three cases stem from the same origin.  The obtained relation is not linear: the slope is the largest for small $a^{-1}$, and then starts leveling off with increasing $a^{-1}$.
A nonlinear relation between $\cos\theta$ and $1/a$ is not new and is sometimes experimentally observed~\cite{drelich1992line, heim2013measurement}, even though approximately linear relations are more commonly reported. Yet, it is important to realize that in typical measurements, droplet radii are varied only within the same order of magnitude~\cite{wang1999line, berg2010impact,mcbride2012influence,leelamanie2012drop, heim2013measurement, zhao2019resolving}. Consequently, variations in the slope, even when they exist, are difficult to discern, especially with noisy data. 

The nonlinear relation does not unequivocally nominate a constant apparent line tension given via \Eq~\ref{eq:cosvs1a}. This inspires us to use the differential definition of the apparent line tension, 
\begin{equation}
\tau\equiv-\gamma \frac{\rmd \cos\theta}{\rmd a^{-1}}
\label{eq:taudef}
\end{equation}
The so-obtained apparent line tension is a function of $a$---it depends on the droplet size.
Using \Eq~\ref{eq:dtheta1} with \Eq~\ref{eq:taudef}, we derive the expression for the apparent line tension caused by surfactants in both deposition scenarios as
\begin{equation}
\tau=\frac{\tau_0}{(1+a^*/a)^2}
\label{eq:tau}
\end{equation}
where we have introduced the asymptotic values%(\ie, for $a\gg a^*$) 
\begin{eqnarray}
\tau_0&=&+\kB T a^*(K_\trm{v}\cos\theta_\trm{w}+K_\trm{s}) c_0\quad\textrm{(scenario A)}\label{eq:tau0I}\\
\tau_0&=&-\kB T a^*(K_\trm{v}\cos\theta_\trm{w}+K_\trm{s}) \tilde c_0\quad\textrm{(scenario B)}\label{eq:tau0II}
\end{eqnarray}
That is, both scenarios are described by the same equations, but with two differences: (i) the opposite sign in $\tau_0$ and (ii) the role of $c_0$ in scenario A is played by $\tilde c_0$ in scenario B.

In \Fig~\ref{fig:theta}c, we plot the apparent line tensions as calculated from \Eqs~\ref{eq:tau} and \ref{eq:tau0I} for the cases in Fig~\ref{fig:theta}a and b (1 mM pentanol in scenario A).
The apparent line tension starts quadratically increasing with base radius as $\tau_0(a/a^*)^2$  (for $a\ll a^*$) and saturates to $\tau_0$ for large base radius (for $a\gg a^*$).

\begin{figure}[h]\begin{center}
\begin{minipage}[b]{0.35\textwidth}\begin{center}
\includegraphics[width=\textwidth]{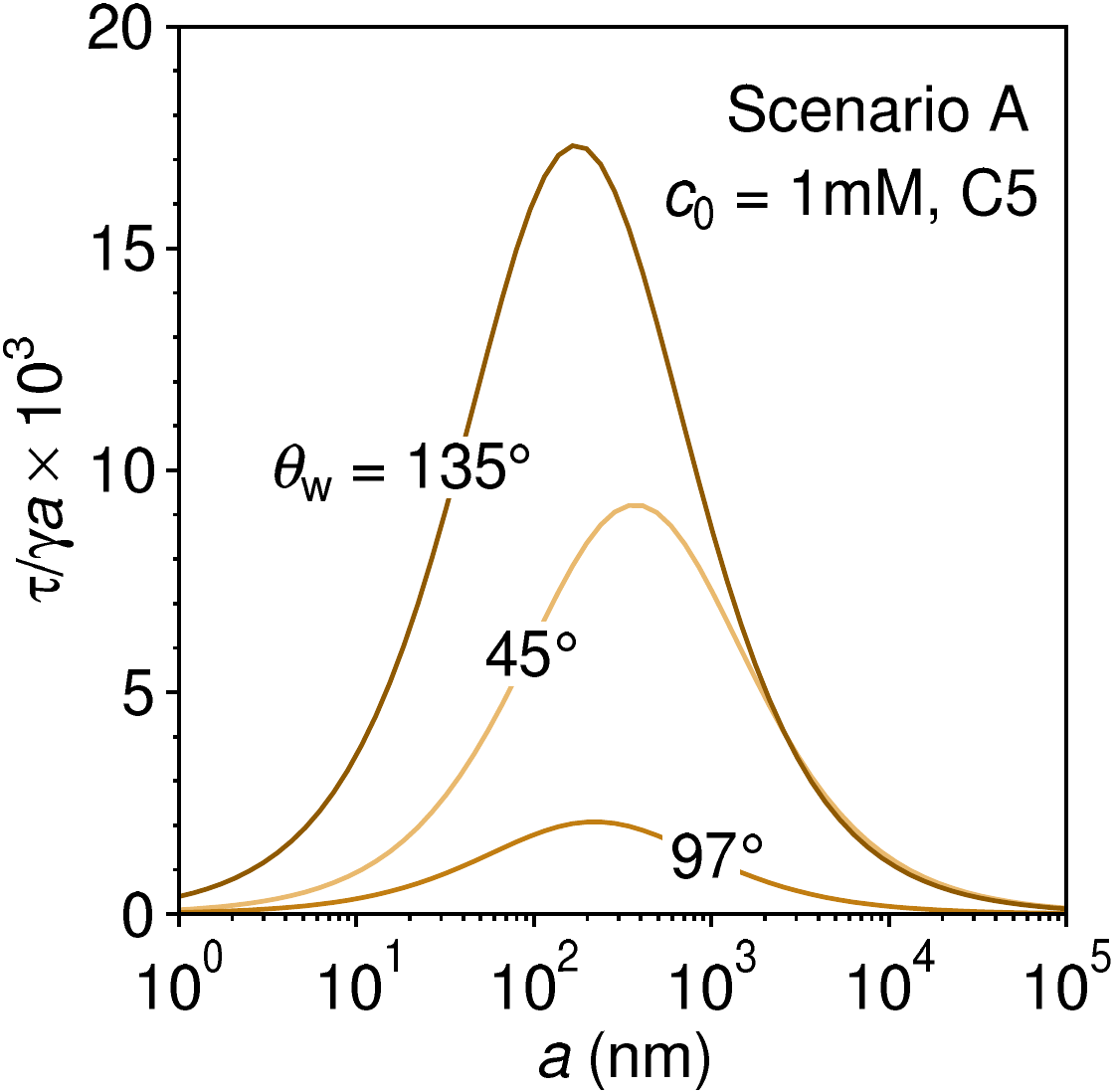}
\end{center}\end{minipage}
\caption{Importance of the apparent line tension to the measurement of $\cos\theta$. Correction to  $\cos\theta$ as a function of droplet size for the case of 1 mM pentanol on three different substrates in scenario A.
}
\label{fig:taulog}
\end{center}\end{figure}

Let us discuss the consequences of the above results.
%The maximal value of the apparent line tension, $\tau_0$, due to surfactants is linearly proportional to their concentrations ($c_0$ in scenario A and $n_A$ in scenario B), and therefore it can reach arbitrarily high values (assuming the linear adsorption regime).
More important than the absolute value of $\tau(a)$ per se is its relative contribution to measurements of size-dependent contact angles.
As suggested by \Eq~\ref{eq:cosvs1a}, the importance of the line tension is given by the quantity $\tau/\gamma a$, which is a finite-size correction to $\cos\theta_0$.
In general, line tension tends to become more important for smaller droplet sizes, but since, in our case, the apparent line tension scales quadratically with small sizes, its importance is nonmonotonic in size. 
As demonstrated in Fig~\ref{fig:taulog}, the correction $\tau/\gamma a$ 
is the largest for intermediate-size droplets and reaches its maximum at $a=a^*$, with the value
\begin{equation}
\left(\frac{\tau}{\gamma a}\right)_\trm{max}=\frac{\tau_0}{4\gamma a^*}%=-\frac{1}{4}\sin\theta\Delta\theta_\trm{w}
\label{eq:max}
\end{equation}
Another significance of the partition radius $a^*$ is that it sets the length scale at which the effect of the apparent line tension due to surfactants is the largest. 

%Recall that $a^*$ is approximately equal to the adsorption coefficients, which scale exponentially with chain length. Thus, size of the surfactant does not only determine the strength of the effect but also segts the lengthscale at wich its influence is most prominent. 

%Finally, the effect is linear in the amount of surfactants. 
%In the next section, we will analyze, at what concentrations of a particular surfactants does the apparent line tension becomes important. 

\begin{figure*}\begin{center}
\begin{minipage}[b]{0.3\textwidth}\begin{center}
\includegraphics[width=\textwidth]{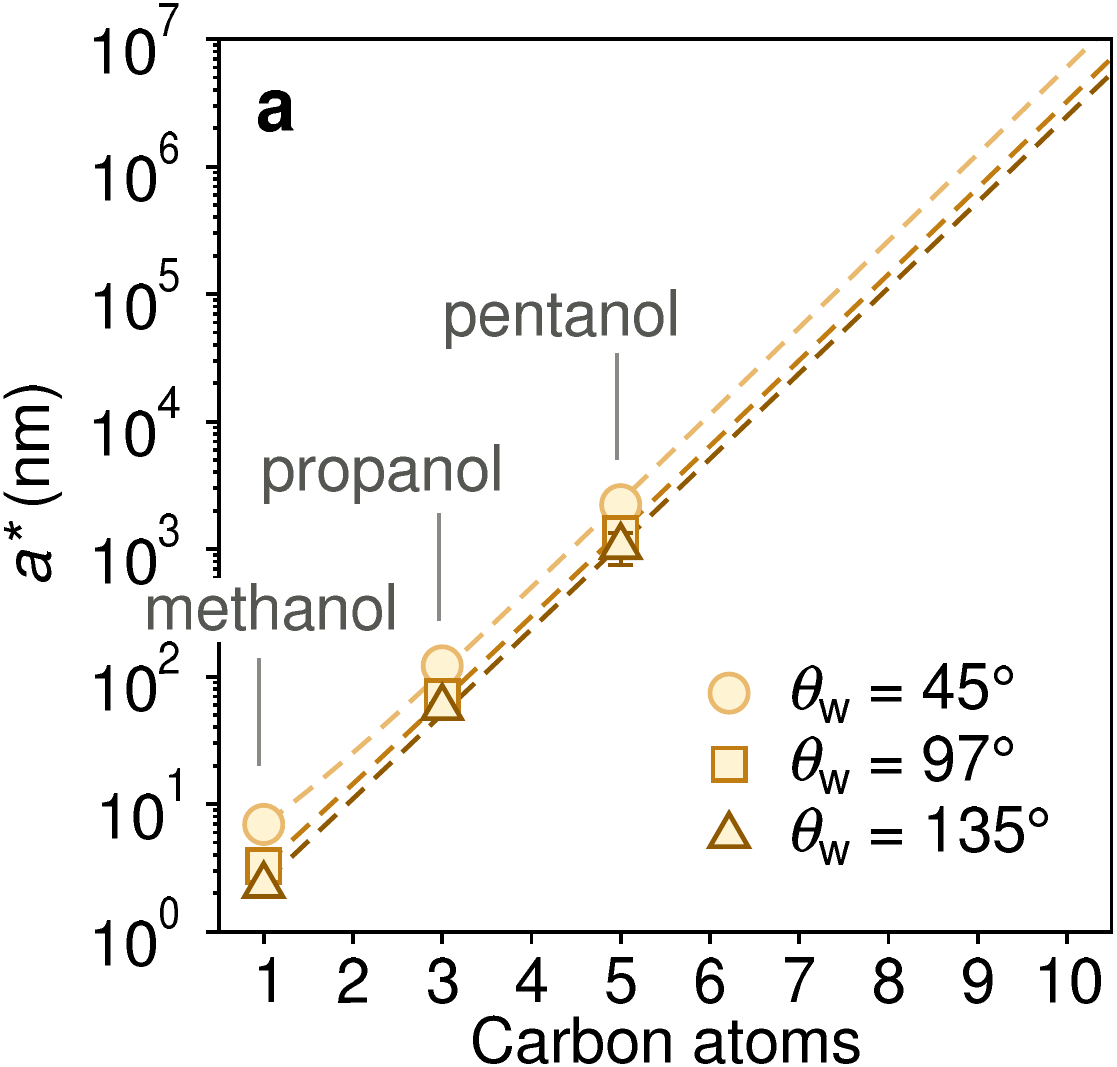} 
\end{center}\end{minipage}\hspace{2ex}
\begin{minipage}[b]{0.3\textwidth}\begin{center}
\includegraphics[width=\textwidth]{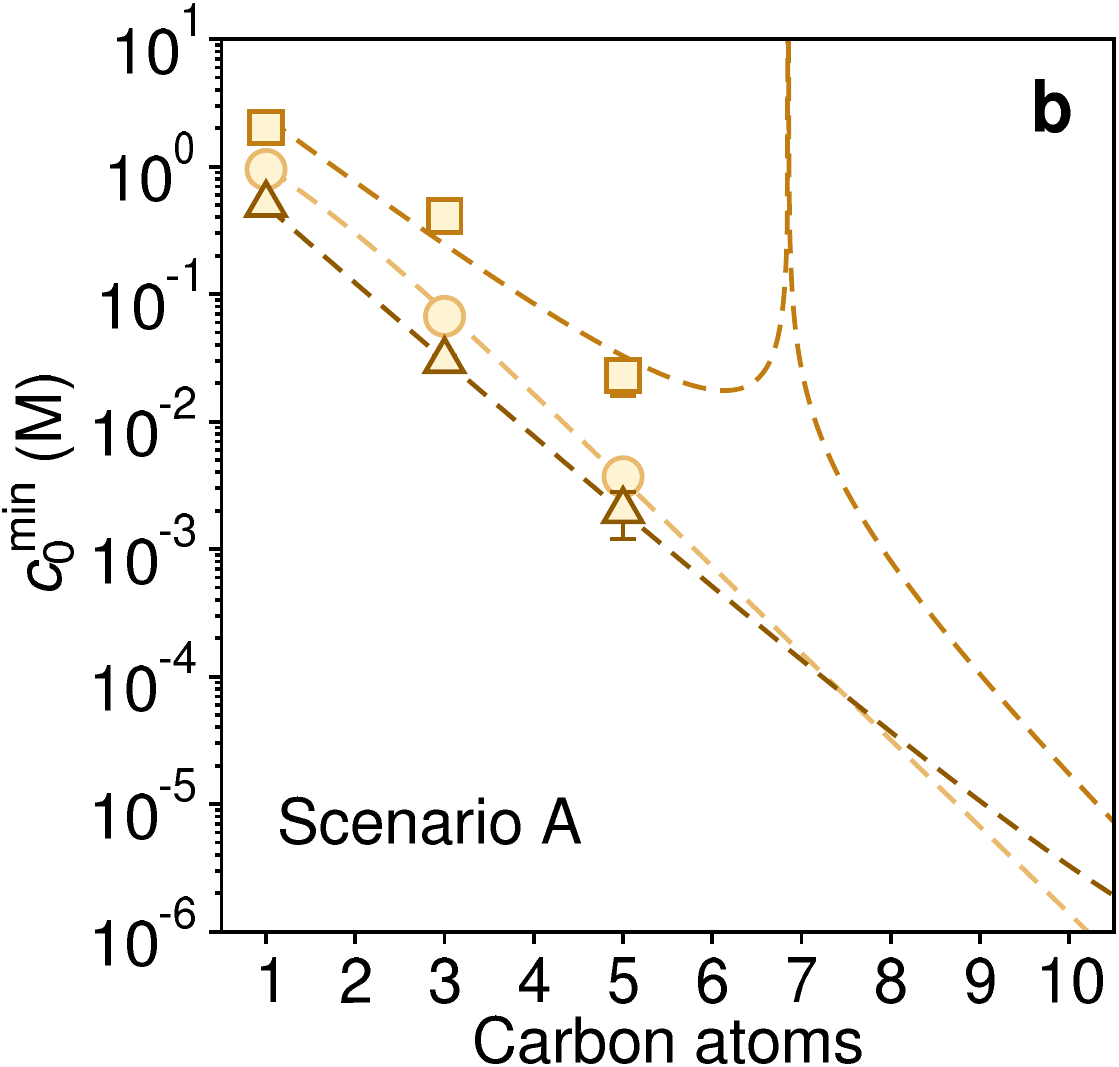}
\end{center}\end{minipage}\hspace{2ex}
\begin{minipage}[b]{0.3\textwidth}\begin{center}
\includegraphics[width=\textwidth]{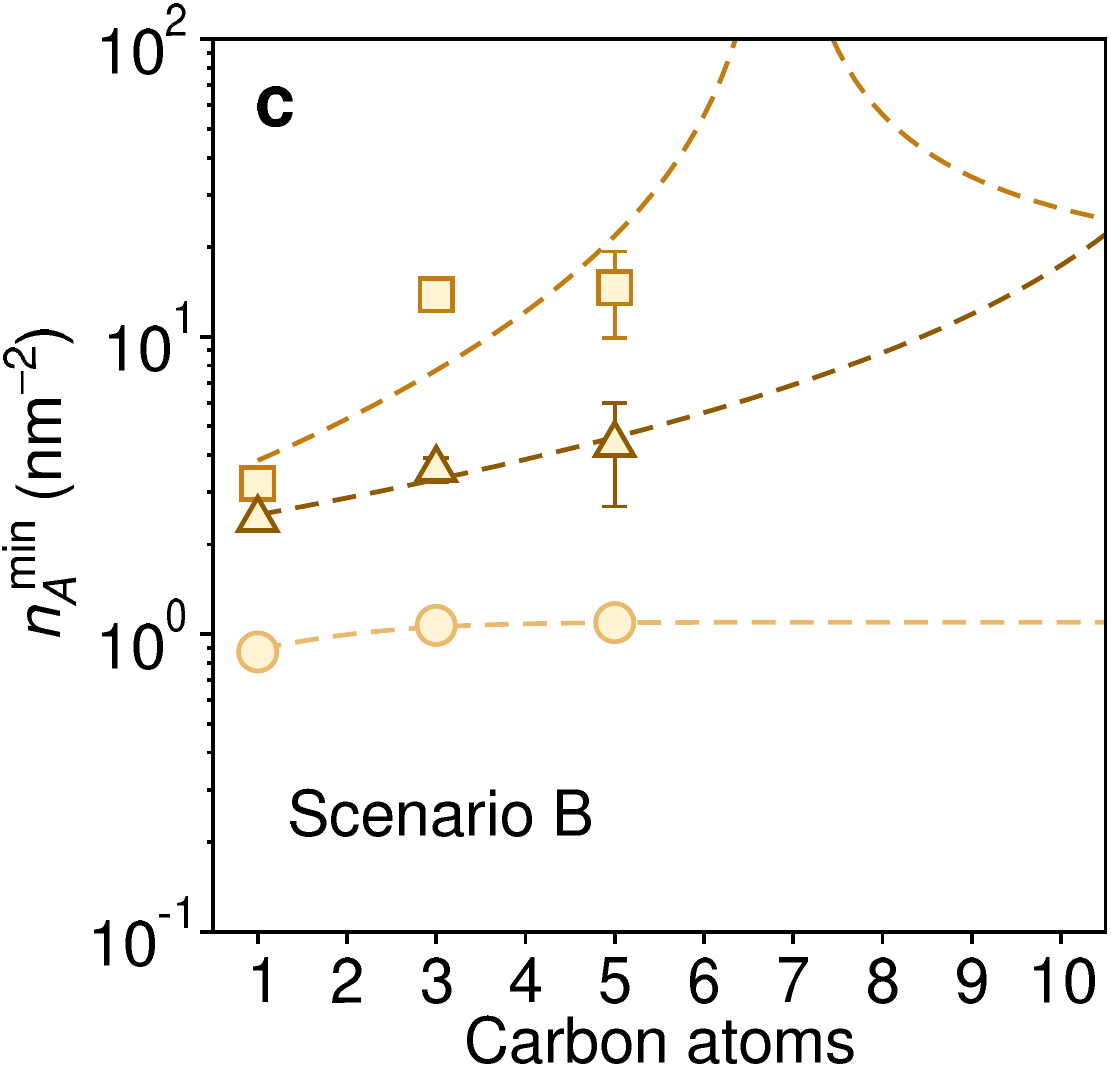}
\end{center}\end{minipage}
\caption{Influence of chain length. (a) Partition radius $a^*$ for linear alcohols as a function of their alkyl chain length. Data are shown for three different substrate contact angles $\theta_\trm{w}$ (the legend applies to all three panels). (b) Minimal bulk concentration that causes a measurable effect in scenario A (\Eq~\ref{eq:minI}). The goniometric uncertainty is chosen as $\delta\theta=1^\circ$. (c) Minimal areal concentration that causes a measurable effect in scenario B (\Eq~\ref{eq:minII}).
The symbols are calculated based on the MD results, and the lines are based on the fitted models for adsorption coefficients (\Eqs~\ref{eq:Kv} and~\ref{eq:Ks}).
}
\label{fig:nC}
\end{center}\end{figure*}

\section{Influence of chain length}%impurities
So far, we have quantified the principles by which low concentrations of surface-active molecules induce an apparent line tension in aqueous droplets and exemplified them with pentanol.
We now extend the discussion to other surfactants in the context of nonionic impurities characterized by different chain lengths.

To that end, we first take a look at how the chain length influences the partition radius $a^*$.
In \Fig~\ref{fig:nC}a, we show $a^*$ as a function of alkyl-tail length for $n$-alcohols. The symbols are the values calculated using the adsorption coefficients for the three alcohols from MD simulations~\cite{staniscia2022tuning}, and the dashed lines are calculated using the fitted models (\Eqs~\ref{eq:Kv} and~\ref{eq:Ks}).
We see that $a^*$ increases approximately exponentially with tail length because $a^*$ is a linear combination of both adsorption coefficients, such that $a^*\sim \max\{K_\trm{v}, K_\trm{s}\}$. With each additional carbon atom in the tail, $a^*$ increases approximately by a factor of 4--5. For the C5 alcohol, $a^*\approx1~\mu$m, whereas for C10, it reaches macroscopic values of $a^*\approx 1$~mm. 

The apparent line tension for droplet sizes of $a\approx a^*$ is approximately $\tau(a^*)\approx \kB T K_\trm{v (s)}^2 c_0\approx \kB T {a^*}^2 c_0$ (scenario A), as long as the linear adsorption regime applies. Because of the quadratic dependence on $a^*$ or adsorption coefficients, each additional carbon atom roughly increases the apparent line tension by a factor of $\sim20$ for a given total concentration, $c_0$. This means that the longer the chain in a surfactant, the lower concentration is needed to cause a detectable contribution to the apparent line tension.
%It then becomes a question, how high concentrations are relevant for contaminants.

This notion motivates us to pose the question: What is the minimal concentration of surface-active impurities $c_0^\trm{min}$ (scenario A) or $n_A^\trm{min}$ (scenario B) needed to cause a measurable effect in the contact-angle variation?
For that, we assume that a goniometric uncertainty in typical contact angle measurements is $\delta \theta=1^\circ\approx 0.017$~rad~\cite{liu2019improving, vuckovac2019uncertainties}.
We then approximate the minimally measurable difference in cosines in \Eq~\ref{eq:cosvs1a} as $\cos\theta-\cos\theta_\trm{w}\approx -\sin\theta_\trm{w}\delta\theta$, which, combined with \Eq~\ref{eq:max} where the effect is the largest, gives us the minimal concentration of impurities that causes a measurable effect in scenario A
\begin{equation}
c_0^\trm{min}=\frac{4\gamma}{\kB T}\frac{\sin\theta_\trm{w}\,\delta\theta}{|K_\trm{v}\cos\theta_\trm{w}+K_\trm{s}|}
\label{eq:minI}
\end{equation}
The value of $c_0^\trm{min}$ can be understood as the detectability threshold, above which the presence of surfactants (e.g., undesired impurities) causes a measurable contribution to the apparent line tension as defined by the modified Young equation (\Eq~\ref{eq:cosvs1a}).

In \Fig~\ref{fig:nC}b, we plot $c_0^\trm{min}$ for the three substrates in scenario A.
First, the threshold concentration decays roughly exponentially with chain length, owing to the fact that $c_0^\trm{min}\propto 1/\max\{K_\trm{v},K_\trm{s}\}$ (see \Eq~\ref{eq:minI}).  
The effect is the smallest on the intermediate substrate with $\theta_\trm{w}\approx 97^\circ$.
For the chain length of around $n_\trm{C}\approx 7$, it happens that $K_\trm{v}\cos\theta_\trm w+ K_\trm{s}\approx 0$. In this particular case, adsorptions of surfactants to both interfaces cancel each other's contribution to the contact angle, as seen from \Eq~\ref{eq:dtheta}, and the droplet's contact angle becomes insensitive to the addition of surfactants.
The effect of surfactants is more pronounced on the hydrophobic and hydrophilic substrates, with an order of magnitude lower detectability threshold.
For the latter two substrates, the calculated threshold value for methanol is slightly below 1~M; for C3, it is 20 mM, whereas for C5, it is around 1 mM. 
Based on our mathematical model (dashed line), the trend continues further with a factor of 4--5 decrease for each added carbon to the hydrophobic chain. For C10, it extrapolates down to $c_0^\trm{min}\approx10^{-6}$~M, which corresponds to the total organic carbon (TOC), a non-specific indicator of water quality, of $\sim 100~\mu$g/L.
This value is 1--2 orders of magnitude below TOC values in tap water and comparable to those of some grades of purified water~\cite{zhang2021critical}. For even longer surfactants, we expect that $c_0^\trm{min}$ decreases even further.
Such low concentrations pose a serious challenge in ensuring a sufficiently clean environment in experimental measurements.
Namely, detecting submicromolar concentrations of surfactant impurities is extremely challenging even with surface-sensitive spectroscopic techniques~\cite{duignan2018detecting, peng2021surface}.
The partition radius of C10 is $a^*\approx 1$~mm, meaning that its effect is most significant for macroscopically-large droplets. 
We calculate that the apparent line tension for $1~\mu$M of C10 is $\tau_0\approx \kB T {a^*}^2 c_0 \approx 10^{-5}$~N. Such high values lie in the top part of the spectrum of reported values in the literature for line tensions, which are also the least understood.
This does not mean, however, that the explanation for such high values lies exclusively in impurities but rather that impurities can be a contributing factor in some cases.

The outcomes of the homologous series of $n$-alcohols, used as model surfactants in this study, are  transferable to other alkyl-based surfactants. 
Many everyday detergents used in industrial and commercial cleaning formulations have alkyl tails of lengths 10 carbons and above.
\del{A prototypical example is sodium dodecyl sulfate (SDS) with the C12 alkyl tail. With an estimated adsorption coefficient at low concentrations of $K_\trm{v}\approx 3\times 10^6$~nm [71, 72],%~\cite{warszynski1998surface, uematsu2018charged},
SDS lies in the ballpark of the $n$-alcohol trend.}
Impurities may consist of a broad mixture of different species that most likely varies from experiment to experiment.
Their contributions could sum up, and polydispersity in the chain length could probably broaden the influential length scale (\Fig~\ref{fig:taulog}), leading to a less size-dependent apparent line tension.

%SDS is a common component of many domestic cleaning, personal hygiene and cosmetic, pharmaceutical, and food products, as well as of industrial and commercial cleaning and product formulations.

Similar to scenario A, we estimate the minimal areal concentration $n_A^\trm{min}$ of surfactants on a pre-contaminated substrate in scenario B that causes a detectable effect in the contact angle variation.
This can be easily derived by replacing $c_0$ in \Eq~\ref{eq:minI} by $\tilde c_0$ (given by \Eq~\ref{eq:ctilde}), which gives us
\begin{equation}
n_A^\trm{min}=\frac{4\gamma}{\kB T}\left(\frac{K_\trm{v}/\cos^2 (\theta_\trm{w}/2)+K_\trm{s}}{K_\trm{v}\cos\theta_\trm{w}+K_\trm{s}}\right)\sin\theta_\trm{w}\,\delta\theta
\label{eq:minII}
\end{equation}
As we show in \Fig~\ref{fig:nC}c, the size of the molecule plays a much less important role in scenario B than in scenario A. 
Here, the adsorption coefficients enter the numerator and the denominator, consequently, the expression in the parentheses is of the order of unity, and the expression becomes almost independent of the surfactant type.
%With this notion, \Eq~\ref{eq:minII} can be further simplified by omitting prefactors to $n_A^\trm{min}\approx \gamma \delta \theta_\trm{w}\sin\theta_\trm{w}/\kB T$, which provides the order of magnitude as $n_A^\trm{min}\approx 1$\cmt{--$10$} nm$^{-2}$.
This is in stark contrast to scenario A, where the surfactant type plays the utmost important role. Thus, in scenario B, \new{the threshold values of $n_A=1$--$10$ nm$^{-2}$ suggest that} nearly the entire surface has to be covered by molecules to cause a measurable effect on apparent line tension. Such circumstances can readily occur when performing experiments in environmental air\new{, especially in urban areas,} because of airborne contamination with various hydrocarbons that quickly cover up the whole surface area~\cite{li2013effect, kozbial2014understanding, smith1998analysis}.
%\new{Nevertheless, sufficient contamination via pre-contaminated surfaces (scenario B) are probably more easily avoidable than in the case of trace pre-existing impurities in bulk liquid (scenario A).}

Our analysis is based on linear Henry's regime of adsorption, which is valid for small enough concentrations. In Appendix~\ref{sec:validity}, we estimate that the range of validity of the linear regime and the subsequent line-tension analysis reaches not far from the detectability threshold. It should be therefore kept in mind that the analysis cannot be quantitatively used for too high surfactant concentrations.
%Prous materials. Leaking of soluble molecules from those pores?

\new{In the end, comparing the outcomes of both scenarios (\Figs~\ref{fig:nC}b and~\ref{fig:nC}c), it appears that trace amounts of pre-existing impurities in the bulk liquid (scenario A) pose a greater challenge to line tension measurements than the contamination of droplets from pre-contaminated surfaces (scenario B).
In the latter case, a cleansed or freshly prepared surface in a clean atmosphere can considerably reduce contamination via this route.}

\section{Conclusions}
%So what?
This study points to the importance of trace amounts of surface-active impurities in measurements of line tension.
We have shown that even tiny amounts of surfactants in deposited aqueous droplets can lead to a notable variation of the contact angles with the droplet's size. A direct consequence is that they contribute to the apparent line tension as traditionally obtained from contact angle measurements of droplets of different sizes. 
%We have shown that even tiny amounts of surface-active impurities in aqueous droplets can notably affect the contact angle variation with droplet size and with that contribute to the apparent line tension of the droplets. We have quantified the phenomenon, which helps estimate the effect's size.
%%%%
We have quantified the phenomenon for a homologous series of $n$-alcohols as a model system for nonionic surfactants and got a good insight into how the effect scales with the alkyl-tail length. %However, the observations are transferable to even more complex organic molecules (notably detergents used for cleaning the equipment, CTAB, SDS).
For longer-tail surfactants, consisting of around 10 carbon atoms and above, already micromolar or submicromolar concentrations cause a measurable effect on the apparent line tension, which can be as large as $10^{-5}$ N.

Such low concentrations highlight the importance of purity in experimental measurements with the aim of reducing contamination.
%Such low concentrations are challenging to detect and pose an important question of purity in experimental measurements and are important to aim to reduce the contamination contribution.
Despite the care taken in the cleanness of lab equipment and using purified water subjected to sophisticated purification techniques, some residual impurities stemming from various sources are unavoidable~\cite{persson2002interfacial,ponce2016effects} and are even challenging to detect~\cite{duignan2018detecting, peng2021surface}.
Therefore, it is essential to realize that these ``background'' impurities limit the experimental resolution of line-tension measurements and, with that, avoid data misinterpretation.
Impurities are thus one of several known factors that can contribute to the measurements of the apparent line tension, but, to the best of our knowledge, hitherto not critically considered as a source of possible artifacts.  
%With this, impurities can contribute to the mystery of hugely scattered line tension measurements, which span over seven orders of magnitude.

It is important to mention that our model was intentionally kept minimal and simple to demonstrate the effect. However, impurities can consist of a broad mixture of different species of different tails and polar-group characters that vary from experiment to experiment. Another viable phenomenon of surfactants is that they can adsorb to the three-phase contact line and, by that, modify the actual line tension directly in the same way as surface adsorption leads to a reduction of the surface tension~\cite{widom2004adsorption}.
\new{Note that our analysis is based on nonionic surfactants. In contrast, ionic surfactants impose a collective electrostatic interplay with counterions and coions, resulting in a more complex and nonlinear adsorption behavior~\cite{uematsu2018charged}.}
These effects, not in the scope of the present paper, are interesting points for future research.
We studied the surfactant effect in water droplets, but qualitatively similar results we expect for other polar liquids with high surface tensions, whereas liquids with lower surface tensions are expected to be less affected by impurities.

In general, trace amounts of surface-active impurities in aqueous solutions often have a negligible effect on bulk properties but can, in some cases, impact the surface behavior enormously. Some other known interface phenomena in which tiny amounts of impurities play visible roles are the Jones--Ray effect~\cite{uematsu2018charged, duignan2018detecting, peng2021surface}, 
negative zeta potentials of hydrophobic surfaces~\cite{zimmermann2001electrokinetic, roger2012hydrophobic,shapovalov2013negligible, uematsu2019impurity}, 
anomalous nanobubble stability~\cite{ducker2009contact, das2010effect},
and some dynamical effects at air-water interfaces~\cite{nicolas2000note, maali2017viscoelastic,arangalage2018dual}.

\appendix 
\section{Change in contact angle}
\label{sec:app:theta}
Here we derive \Eq~\ref{eq:dtheta1}, which gives the change in the contact angle due to surfactants. 
Inserting \Eqs~\ref{eq:Gammawv}--\ref{eq:Gammasv} into \Eq~\ref{eq:LR} and performing the derivation of the left-hand side leads to
\begin{equation}
\cos\theta-\gamma\sin\theta\,\frac{\rmd\theta}{\rmd c}\frac{\rmd c}{\rmd \gamma}=-\frac{K_\trm{s}}{K_\trm{v}}
\label{eq:App1}
\end{equation}
We have used the chain rule to express the derivative of $\theta$ with respect to bulk concentration $c$. Next, the water--vapor surface tension in the linear adsorption regime changes with adsorption as $\rmd\gamma=-\kB T \rmd \Gamma_\trm{wv}$, which further leads to $\rmd\gamma/\rmd c = -\kB T K_\trm{v}$.
Inserting the last expression into \Eq~\ref{eq:App1} and after some rearrangement, we obtain
\begin{equation}
\frac{\gamma\sin\theta\,\rmd \theta}{K_\trm{v}\cos\theta+K_\trm{s}}= -\kB T \rmd c
\end{equation}
After the integration from $\theta_\trm{w}$ to $\theta$ and $0$ to $c$ of the left- and right-hand side, respectively, we get
\begin{equation}
\frac{\gamma}{K_\trm{v}} \,\ln\frac{K_\trm{v}\cos\theta+K_\trm{s}}{K_\trm{v}\cos\theta_\trm{w}+K_\trm{s}} = \kB T c
\end{equation}
Another small rearrangement gives \Eq~\ref{eq:dtheta1}.

\section{Spherical cap geometry}
\label{sec:app:cap}
A sessile droplet deposited on a solid flat substrate assumes the form of a spherical cap with the contact angle $\theta$ and base radius $a$. Its volume is then
\begin{equation}
V=\frac{\pi a^3}{3}\,\frac{(2+\cos\theta)\sin\theta}{(1+\cos\theta)^2}
\label{eq:Vdeposited}
\end{equation}
the water--vapor surface area (the cap) is
\begin{equation}
A_\trm{cap}=\frac{2\pi a^2}{1+\cos\theta}
\end{equation}
and the surface area of its base is
\begin{equation}
A_\trm{base}=\pi a^2
\end{equation}

\section{Regime of validity}
\label{sec:validity}

\new{The bottleneck of the linear adsorption regime (\Eqs~\ref{eq:Gammawv} and \ref{eq:Gammasw}) is that long-tailed surfactants could considerably interact with each other already at relatively low adsorption densities. The onset of these interactions can be estimated on the second-order virial expansion level, which gives the surface-tension decrement as~\cite{staniscia2022tuning}
\begin{equation}
\Delta \gamma = -\kB TK_\trm{v} c + \kB T B_2 K_\trm{v}^2 c^2
\label{eq:B2}
\end{equation}
Here, $B_2$ is the two-dimensional second virial coefficient of surfactants at the interface.
The interaction between surfactants at an interface depends on the balance between the hydrophilic and hydrophobic groups and can be best explored from surface-pressure isotherms. 
For surfactants that sterically repel, $B_2$ is positive and roughly equal to the surfactant's cross-section area, $B_2\simeq A_\trm{surf}$. } Based on \Eq~\ref{eq:B2}, this defines the range of the linear regime as $c_0^\trm{lin}=(K_\trm{v} A_\trm{m})^{-1}$.
The order of magnitude of the ratio between $c_0^\trm{min}$ (\Eq~\ref{eq:minI}) and $c_0^\trm{lin}$, assuming $|K_\trm{v}\cos\theta_\trm{w}+K_\trm{s}|\sim K_\trm{v}$ and $\sin\theta_\trm{w}\sim 1$, is
\begin{equation}
\frac{c_0^\trm{min}}{c_0^\trm{lin}}\sim \frac{\gamma A_\trm{m}}{\kB T}\,\delta\theta
\end{equation}
which for $A_\trm{surf}\sim 0.2$--$1.0$ nm$^2$ gives ${c_0^\trm{min}}/{c_0^\trm{lin}}\approx 0.05$--$0.2$. Hence, the detectability threshold lies within the linear regime.
%For higher concentrations, the non-linearity in adsorption models should be take into account.

\new{On the contrary, for attractive surfactants that tend to cluster, $B_2$ can be large and negative, such that the linearity can break down already at lower concentrations. 
As seen from eq~\ref{eq:B2}, the surface tension is reduced more for attractive surfactants (featuring $B_2<0$) at a given concentration compared to repulsive surfactants (featuring $B_2>0$).
From this, we qualitatively conclude that strongly attractive surfactants impose an even more 
substantial influence on the apparent line tension.
}

A similar estimate can be done for scenario B and repulsive surfactants. Assuming $|K_\trm{v}/\cos^2(\theta_\trm{w}/2)+K_\trm{s}|\sim K_\trm{v}$, we obtain
\begin{equation}
n_A^\trm{lin}\approx \frac{2}{A_\trm{surf}}
\end{equation}
which for $A_\trm{surf}\sim 0.2$--$1.0$ nm$^2$ gives $n_A^\trm{lin}\approx 2$--$10$~nm$^{-2}$, which is right at the crossover of the linear regime.

\section*{Acknowledgments}
We acknowledge financial support from the Slovenian Research Agency (contracts P1-0055 and J1-1701). 

\section*{Conflict of Interest} 
The authors have no conflicts to disclose.

\section*{Author Contributions} 
{\bf Fabio Staniscia:} Formal analysis (supporting); Investigation (equal); Writing – original draft (supporting).
{\bf Matej Kandu\v{c}:} Conceptualization (lead); Formal analysis (lead); Investigation (equal); Funding acquisition (lead); Visualization (lead); Writing – original draft (lead).

%\subsection*{Supporting Information Description}
%TI of an ion pair; Potential across a planar interface; Partitioning of ions; Ion distribution across the interface: mean-field level

\footnotesize
\setlength{\bibsep}{0pt}
\bibliography{main}

\end{document}